%% file: transRec.tex
\documentclass[conference]{IEEEtran}
\IEEEoverridecommandlockouts

\usepackage{amssymb, multirow}

\input{preamble}

\begin{document}

\title{Cross-Domain Collaborative Filtering via Translation-based Learning}

\author{\IEEEauthorblockN{Dimitrios Rafailidis}
\IEEEauthorblockA{\textit{Maastricht University} \\
Maastricht, The Netherlands \\
dimitrios.rafailidis@maastrichtuniversity.nl}
}

\maketitle

\begin{abstract}
With the proliferation of social media platforms and e-commerce sites, several cross-domain collaborative filtering strategies have been recently introduced to transfer the knowledge of user preferences across domains. The main challenge of cross-domain recommendation is to weigh and learn users' different behaviors in multiple domains. In this paper, we propose a Cross-Domain collaborative filtering model following a Translation-based strategy, namely CDT. In our model, we learn  the embedding space with translation vectors and capture high-order feature interactions in users' multiple preferences across domains. In doing so, we efficiently compute the transitivity between feature latent embeddings, that is if feature pairs have high interaction weights in the latent space, then feature embeddings with no observed interactions across the domains will be closely related as well. We formulate our objective function as a ranking problem in factorization machines and learn the model's parameters via gradient descent. In addition, to better capture the non-linearity in user preferences across domains we extend the proposed CDT model by using a deep learning strategy, namely DeepCDT. Our experiments on six publicly available cross-domain tasks demonstrate the effectiveness of the proposed models, outperforming other state-of-the-art cross-domain strategies.
\end{abstract}


\section{Introduction}
The collaborative filtering strategy has been widely followed in recommendation systems, where users with similar preferences tend to get similar recommendations. User preferences are expressed explicitly in the form of ratings or implicitly in the form of number of views, clicks, purchases, and so on. Representative collaborative filtering strategies are latent models such as Matrix Factorization and Factorization Machines (FMs)~\cite{Rendle10}, which factorize the data matrix with user preferences in a single domain (e.g., music or video), to reveal the latent associations between users and items. However, data sparsity and cold-start problems degrade the recommendation accuracy, as there are only a few preferences on which to base the recommendations in a single domain. With the advent of social media platforms and e-commerce systems, such as Amazon and Netflix, users express their preferences in multiple domains. For example, in Amazon users can rate items from different domains, such as books and retail products, or users express their opinion on different social media platforms, such as Facebook and Twitter. In the effort to overcome the data sparsity and cold-start problems, several cross-domain recommendation strategies have been proposed, which exploit the additional information of user preferences in multiple auxiliary/source domains to leverage the recommendation accuracy in a target domain~\cite{AliannejadiRC17, LI09, Lon14, RafailidisC16, RafailidisC17}. However a pressing challenge resides on how to transfer the knowledge of user preferences from different domains, by also weighting the importance of users' different behaviors accordingly.

In cross-domain recommendation, the source domains can be categorized based on users' and items' overlaps, that is, full-overlap, and partial or non user/item overlap between the domains. In this study, we focus on partial users' overlaps between the target and the source domains, as it reflects on the real-world setting~\cite{Hu18}. Cross-domain recommendation algorithms differ in how the knowledge of user preferences from the source domains is exploited, when generating the recommendations in the target domain. For example,  Li et al.~\cite{LI09} calculate user and item clusters for each domain, and then encode the cluster-based patterns in a shared codebook; finally, the knowledge of user preferences is transferred across domains through the shared codebook. Gao et al.~\cite{GAO13} present a Cluster-based Latent Factor Model which uses joint nonnegative tri-factorization to construct a latent space to represent the rating patterns of user clusters on the item clusters from each domain, and then generates the cross-domain recommendations based on a subspace learning strategy. Cross-Domain collaborative filtering with FMs, presented in~\cite{Lon14}, is a state-of-the-art cross-domain recommendation. It is a context-aware approach which applies factorization on the merged domains, aligned by the shared users, where the source domains are used as context. Hu et al.~\cite{Hu18} jointly learn neural networks to generate cross-domain recommendations based on stich units, introducing a shared auxiliary matrix to couple two hidden layers when training the networks in parallel. However, these cross-domain recommendation strategies do not pay attention to users' complex behaviors across domains, where we have to weigh and learn high-order feature interactions of users' preferences  while transferring the knowledge of users' selections from multiple domains.

Recently, metric/translation-based models have been shown to be effective in collaborative filtering~\cite{HsiehYCLBE17}. Essentially the idea behind is to learn an embedding and translation space for each feature dimension, replacing the inner product of latent models with the squared Euclidean distance to measure the interaction strength between features. By learning a latent item embedding space with translation vectors, such models can better capture the transitivity between feature latent embeddings also known as the similarity propagation process, that is if feature pairs have high interaction weights in the latent space, then feature embeddings with no observed interactions will be closely related as well. In a similar spirit, recent studies demonstrate the effectiveness of translation vectors in sequential recommendation~\cite{PasrichaM18,HeKM17}. Nonetheless, these  models produce recommendations in a single-domain, and omit users' various and complex feature interactions across domains.

To overcome the shortcomings of existing strategies, we propose a Cross-Domain Translation-based learning model, namely CDT, making the following contributions: We first compute the cluster-based users' similarities across domains, and then we propose a metric/translation-based learning strategy to compute the embedding space with translation vectors and capture high-order feature interactions when optimizing user preferences across domains. We formulate our objective function as a ranking problem in FMs and learn the model's parameters via gradient descent. In addition, to efficiently learn the non-linearity in user preferences across domains we extend the proposed CDT model by using a deep learning strategy, namely DeepCDT. In our experiments on six cross-domain tasks, we show that our models outperform other single-domain and cross-domain strategies.

\section{The Proposed Model} \label{sec:prop}
In our setting we assume that we have $d$ different domains, where $n_p$ and $m_p$ are the numbers of users and items in the $p$-th domain, respectively. In matrix $R^{(p)}$, we store the user preferences on items, in the form of explicit feedback e.g., ratings or in the form of implicit feedback e.g., number of views, clicks, and so on. In this study we consider users' partial overlaps across the domains. We define a users' overlapping matrix $O^{(pt)} \in \mathbb{R}^{n_p \times n_t}$ between a source domain $p$ and the target domain $t$. For each cell holds $O^{(pt)}(k,u)$= 1, if users $k$ and $u$ are the same user in domains $p$ and $t$, and 0 otherwise. The goal of the proposed CDT model is \emph{to generate personalized recommendations in the target domain $t$, while transferring and weighting users' different preferences from the $d-1$ source domains}. The proposed CDT model consists of a cross-domain co-clustering strategy to compute the cluster-based similarities among different users across domains, and our translation-based strategy to generate cross-domain recommendations.

\subsection{Cross-domain Co-Clustering} \label{sec:clust}
In each domain we first compute the user cluster assignment matrices $C^{(p)}\in \mathbb{R}^{n_p \times c_p}$ and $C^{(t)}\in \mathbb{R}^{n_t \times c_t}$, where $c_p$ and $c_t$ are the number of user clusters in the $p$-th domain and the target domain $t$, respectively. In our implementation we compute matrices $C^{(p)}$, with $p=1,\ldots,d-1$, and matrix $C^{(t)}$ based on the graph Laplacian method~\cite{LAP13}. To calculate the cluster-based similarities of users in matrix $Y^{(pt)} \in \mathbb{R} ^{c_p \times c_t}$ between domains $p$ and $t$, we follow a co-clustering strategy for each source domain $p$ and the target domain $t$, trying to minimize the following objective function:
\begin{equation}\label{eq:obj2}
\begin{array}{c}
\min\limits_{Y^{(pt)}} ||O^{(pt)} - C^{(p)} Y^{(pt)} {C^{(t)}}^\top||_F^2 +\lambda||Y^{(pt)}||_{2,1} \\
\text{subject to ${Y^{(pt)}}^\top Y^{(pt)}=I$, $Y^{(pt)}\geq 0$}
\end{array}
\end{equation}
$||\cdot||_{2,1}$ denotes the $L_{2,1}$ norm of a matrix forcing $Y^{(pt)}$ to be sparse, reflecting on the real-world scenario, where users' overlaps are usually sparse~\cite{CREM11}. Then, to compute the common user's $u$ behavior in domains $p$ and $t$ we calculate each cell of matrix $Q^{(pt)} \in \mathbb{R}^{m_p \times n_t}$ based on the entries of the cluster-based user similarities in $Y^{(pt)}$ as follows:
\begin{equation} \label{eq:pame}
Q^{(pt)}(h,u) = \frac{\sum\limits_{k \in \mathcal{I}^{(p)}_h} Y^{(pt)}\big(C^{(p)}(k),C^{(t)}(u)\big)}{\big|\mathcal{I}^{(p)}_h \big|}
\end{equation} 
where $\mathcal{I}^{(p)}_h$ is the set of users $k=1,\ldots,n_p$ that have interacted with item $h=1,\ldots,m_p$ in the $p$-th domain, and $(k,h) \in R^{(p)}$.

\subsection{Cross-domain Translation-based Learning} \label{sec:eff}
\subsubsection{Feature Vector} For the target domain $t$ we represent each user-item interaction $(u, i) \in R^{(t)}$ as a feature vector $f \in \mathbb{R}^{n_t+m_t}$, with $u=1,\dots,n_t$ and $i=1,\ldots,m_t$. Using a lookup operation, the feature vector is expressed by its sparse representation as follows: $f^{(t)}=\big\{(u,t), (i,t) \big\}$. For each source domain $p=1,\dots,d-1$, we extend $f^{(t)}$ with a vector  $z^{(p)}(u)=\big\{h, Q^{(pt)} (h,u)  \big\}$,  with $h=1,\ldots,m_p$ and $(u,h) \in R^{(p)}$. When computing the vector $z^{(p)}(u)$, matrix $Q^{(pt)}$ weighs the ratings on $h$ of user $u$ in the $p$-th domain. By augmenting the feature vector $f^{(t)}$ with the $d-1$ vectors $z^{(p)}(u)$ of the source domains, its sparse representation becomes an $l$-dimensional vector as follows:
\begin{equation}
x^{(t)} = \big\{\underbrace{(u,t), (i,t),}_{\text{target knowledge}} \underbrace{z^{(1)}(u),\ldots, z^{(d-1)}(u)} _{\text{source knowledge}} \big\}
\end{equation}
Notice that vector $x^{(t)}$ contains the user-item interactions in domain $t$, as well as the interactions in $p$, weighted by $Q^{(pt)}$ based on Eq.~(\ref{eq:pame}).

\subsubsection{Embedding and Translation Vectors}\label{sec:emb} For each dimension of $x^{(t)}$ we try to learn the embedding vectors $v_i \in \mathbb{R}^q$, $v_h \in \mathbb{R}^q$ and a translation vector $v'_i \in \mathbb{R}^q$, where $q$ is the number of latent dimensions,  $i\in\mathcal{I}^{(t)}$ is an item that belongs to the item set that a user $u$ has rated in the target domain $t$ and $h\in\mathcal{I}^{(p)}$ in the source domain $p$. To capture the user's $u$ transition from item $i$ in domain $t$ to item $h$ in domain $p$, we have to learn a translation vector as follows: $v_i + v'_i \approx v_h$. This means that $v_h$ should be a nearest neighbor of $v_i + v'_i$ in the $q$-th dimensional latent space according to a metric-based metric e.g., the squared Euclidean distance $d^2(v_i+v'_i,v_h)$. Notice that the distance $d^2(v_i+v'_i,v_j)$ replaces the inner product term of conventional latent models, following the metric/translation-based learning strategy~\cite{HsiehYCLBE17}. In doing so, the proposed CDT model can effectively capture the transitive property between feature embeddings of domains $t$ and $p$ in the the latent space. As a consequence, in our cross-domain model if feature pairs have high interaction weights in the latent space, then feature embeddings with no observed interactions across the domains will be closely related as well.  Based on the collaborative filtering strategy of FMs~\cite{Rendle10}, we can formulate the model Equation of CDT as follows:
\begin{equation}\label{eq:mod}
\hat{y}(x^{(t)}) = w_0 + \sum\limits_{i=1}^{l}w_i x^{(t)}_i + \sum\limits_{h=i+1}^{l} d^2(v_i + v'_i,v_h) x^{(t)}_i x^{(t)}_h
\end{equation}
$w_0$ is the global bias, $w_i$ is the linear term for feature $x^{(t)}_i$.  Accordingly, $v_i$ and $v'_i$ are the embedding and translation vectors for feature $x^{(t)}_i$.

\subsubsection{Objective Function}
Following the Bayesian Personalized Ranking criterion~\cite{BPR}, we try to rank higher the observed items $i\in\mathcal{I}^{(t)}$ than the unobserved ones $j\in\mathcal{I}'^{(t)}$ for a user $u\in \mathcal{U}^{(t)}$ in the target domain $t$. In addition, in our cross-domain setting we have to rank higher the observed items $h\in\mathcal{I}^{(p)}$ in a source domain $p$ than the unobserved ones $j\in\mathcal{I}'^{(t)}$ in the target domain $t$. To describe these relations we first consider a global item set $\mathcal{I}_u$ that user $u$ has rated in all the $d$ domains. Then, we define the partial relation $i' >_u j$, with $i' \in \mathcal{I}_u$. In our model, we formulate the following objective function:

\begin{equation}\label{eq:loss1}
\hat{\Theta} = \arg\max_{\Theta} = \prod_{u \in {\mathcal{U}^{(t)}}} \prod_{i' \in \mathcal{I}_u}  \prod_{j\in\mathcal{I}'^{(t)}} Pr(i' >_u j|\Theta)Pr(\Theta)
\end{equation}
with $\Theta$ being the set of the model parameters. According to the model Equation of CDT in~Eq.~(\ref{eq:mod}), the objective function of Eq.~(\ref{eq:loss1}) is formulated as follows:
\begin{equation}\label{eq:loss2}
\hat{\Theta} = \arg\max_{\Theta} = \prod_{u \in {\mathcal{U}^{(t)}}} \prod_{i' \in \mathcal{I}_u}  \prod_{j\in\mathcal{I}'^{(t)}} \ln\sigma\big( \hat{y}( x^{(t)}_{u,i'} ) - \hat{y}(  x^{(t)}_{u,j} ) \big) - \Omega(\Theta)
\end{equation}
where $\sigma$ the sigmoid function and $\Omega(\Theta)$ is the $L_2$-norm regularization term on the model's parameters. To compute the model's parameters we transform the objective function of Eq~(\ref{eq:loss2}) as a minimization problem and optimize it via gradient descent using negative sampling, that is randomly selecting unobserved items in the target domain $t$. In our implementation, we used five negative samples for each positive/observed sample in the target domain $t$, as we found out that for larger numbers of negative samples the computational cost of the model learning did not pay off in terms of recommendation accuracy

\subsubsection{DeepCDT} As the user-item interactions are non-linearly associated across the domains, we implemented a variant of the proposed CDT model, namely DeepCDT. Using the model Equation of CDT in~Eq.~(\ref{eq:mod}) as the bottom layer of a deep neural network with $e$ hidden layers, we learn the deep representations of the translated-based features of Section~\ref{sec:emb}, in a similar way as the single-domain deep learning strategy of FMs in~\cite{ZhangDW16}. In our implementation we used Tensorflow\footnote{\url{https://www.tensorflow.org}} and fix the number of hidden layers $e$=5. We computed the model's parameters, that is the weight matrices of the neural network via backpropagation with stochastic gradient descent, trying to optimize the objective function in Eq.~(\ref{eq:loss2}). We employed mini-batch Adam which adapts the learning rate for each parameter by performing smaller updates for frequent and larger updates for infrequent parameters. We set the batch size of mini-batch Adam to 512 with a learning rate of 1e-4. Also, we varied the number of latent dimensions $q$ from 10 to 100 by a step of 10, using a grid selection strategy and we kept the latent dimensions fixed based on cross-validation.

\section{Experiments} \label{sec:exp}
\subsection{Setup}
\subsubsection{Cross-domain Tasks} Our experiments were performed on six cross-domain tasks from the Amazon dataset~\cite{LeskovecAH07}. The items are grouped in categories/domains, and we evaluate the performance of our model on the six largest domains. The main characteristics of the evaluation data are presented in Table~\ref{tab:data}. 

\begin{table}[h]
\centering
\caption{The six cross-domain recommendation tasks.} \label{tab:data}
\vspace{-0.2cm}
\begin{center}%
\begin{tabular}{l@{\quad\quad}c@{\quad}c@{\quad}c}\hline
Domain & Users & Items & Ratings  \\ \hline
Electronics & 18,649 & 3,975  &  23,009\\
Kitchen & 16,114 & 5,511  & 19,856 \\
Toys & 9,924 & 3,451 & 13,147 \\
DVD & 49,151 & 14,608 & 124,438 \\
Music & 69,409 & 24,159 & 174,180 \\
Video & 11,569  &5,223  & 36,180 \\ \hline
\end{tabular}
\end{center}
\end{table}

\subsubsection{Evaluation Protocol} In each out of the six cross-domain recommendation tasks, the goal is to generate recommendations for a target domain, while the remaining five domains are considered as source domains. We trained the examined models on the 50\% of the target domain and all the ratings of the source domains as training set. We used 10\% of the ratings in the target domain as cross-validation set to tune the models' parameters and evaluate the examined models on the remaining test ratings. To remove user rating bias from our results, we considered an item as relevant if a user has rated it above her average ratings and irrelevant otherwise~\cite{Rec17}. We measured the quality of the top-$n$ recommendations in terms of the ranking-based metrics recall and Normalized Discounted Cumulative Gain (NDCG@n). Recall is the ratio of the relevant items in the top-$n$ ranked list over all the relevant items for each user. NDCG measures the ranking of the relevant items in the top-$n$ list. For each user the Discounted Cumulative Gain (DCG) is defined as: $DCG@n = \sum_{j=1}^{n}{\frac{2^{rel_j}-1}{\log_2{j+1}}}$, where $rel_j$ represents the relevance score of item $j$, that is binary in our case, i.e., relevant or irrelevant. NDCG is the ratio of DCG/iDCG, where iDCG is the ideal DCG value given the ratings in the test set. We fixed the number of recommendations to $n$=10. We repeated our experiments five times and averaged recall and NDCG over the five runs. 

\subsubsection{Compared Methods} We compare the proposed models CDT and DeepCDT with the single-domain strategies FMs~\cite{Rendle10}, CML~\cite{HsiehYCLBE17} and FNN~\cite{ZhangDW16}, and the cross-domain strategies CBT~\cite{LI09}, CLFM~\cite{GAO13}, CDCF~\cite{Lon14} and ScoNet~\cite{Hu18}. The parameters of the examined methods have been determined via cross validation and in our experiments we report the best results. 

\begin{table}[t]
\centering
\caption{Effect on recall. Bold values denote the best scores, using the paired t-test ($p<$0.05). The underlined values denote the second best method. } \label{tab:res1}
\vspace{-0.2cm}
\begin{center}%
\resizebox{\columnwidth}{!}{
\begin{tabular}{l@{\quad}c@{\quad}c@{\quad}c@{\quad}c@{\quad}c@{\quad}c}\hline
			 	& Electronics  & Kitchen & Toys & DVD & Music & Video\\\hline
FMs 			&	.183 &	 .124 &	 .138 & .462 	& .431	 & .601	  \\
CML 			&	.201  &	 .139 & .142 &	 .488 &	 .429 &	 .629  \\
FNN			&	.212	 &	.136 & 	.146 &	.502 &	.447 &	.613  \\
CBT			&	.221  &	.142 &	.157 &	.519 &	.451 &	.653  \\
CLFM		&	.239  &	.151 &	.166 &	.534 &	.458 &	.672  \\
CDCF 		&	.234  &	.155 &	.171 &	.568 &	.462 &	.684  \\
ScoNet 	&	\underline{.248}  &	.173 & .176 & \underline{.605}	 &	\underline{.493} & .697	  \\  
CDT 			&	.241	 &	\underline{.178} & \underline{.185}	 &	 .583 &	.488 &	\underline{.706}  \\
DeepCDT  &	 \textbf{.265 } & \textbf{.194}	& \textbf{.204} & \textbf{.634}	 &	\textbf{.524} &	\textbf{.713}  \\ \hline
\end{tabular}}
\end{center}
\end{table}

\subsection{Performance Evaluation}
Tables~\ref{tab:res1} presents the experimental results in terms of recall. In addition, Figure 1 shows the effect on NDCG for the cross-domain models in the largest domain ``Music'', by varying the training set size. The cross-domain models CBT, CLFM, CDCF, ScoNet, CDT and DeepCDT significantly outperform the single-domains models of FMs CML and FNN, by exploiting users' preferences in the source domains when generating recommendations, thus reducing the data sparsity in the target domain. The proposed CDT model clearly beats the baseline cross-domain strategies CBT and CLFM, and the cross-domain strategy CDCF with FMs, demonstrating that our translation-based strategy plays a crucial role in boosting the cross-domain recommendation accuracy. To further verify the importance of our translation-based strategy, we observe that DeepCDT achieves higher performance than FNN. Notice that instead of feeding the neural network with the features of FMs from a single-domain as FNN~\cite{ZhangDW16}, in our cross-domain setting we feed the neural network with the translated-based features, learned in Section~\ref{sec:emb}, which explains DeepCDT high recommendation accuracy. ScoNet and CDT have comparable performance because ScoNet captures the non-linear associations between user preferences across the domains, while CDT only models the high-order feature interactions in users' behaviors between the domains. Using the paired-t test we found out that DeepCDT is superior over all the competitive approaches for $p<$0.05, by taking into account both the non-linearity and  high-order feature interactions in users' behaviors across multiple domains. 

\section{Conclusions} \label{sec:conc}
We presented CDT, a translation-based model for generating cross-domain recommendations. The key idea of our CDT model is to transfer the knowledge of users' preferences across domains, while computing the translation-based feature interactions, and consequently capturing complex high-order feature interactions in different domains. In addition, we introduce a deep learning variant of our model, namely DeepCDT to efficiently compute the non-linear associations of features in users' behaviors across the domains. Our experiments showed that the proposed approaches significantly outperform other baseline methods, proving the importance of our cross-domain translation-based learning strategy.

As future work, we plan to extend the proposed CDT and DeepCDT models for sequential recommendations in cross-domain tasks. Generating sequential recommendations is a challening task, where the goal is to predict the next item that a user will select. Although there are many single-domain strategies for sequential recommendation, such as the studies reported in~\cite{PasrichaM18,HeKM17}, the case of cross-domain reflects better on the real-world scenario, where not only we have to capture users' sequential behaviors, but transfer this knowledge across different domains. In addition, we plan to study the influence of the proposed method on the link prediction task across multiple platforms~\cite{AntarisRN14}, as well as on modeling users' preference dynamics~\cite{RafailidisN15a}. 

\begin{figure}[t] \centering
\includegraphics[width=8.5cm,height=5.7cm]{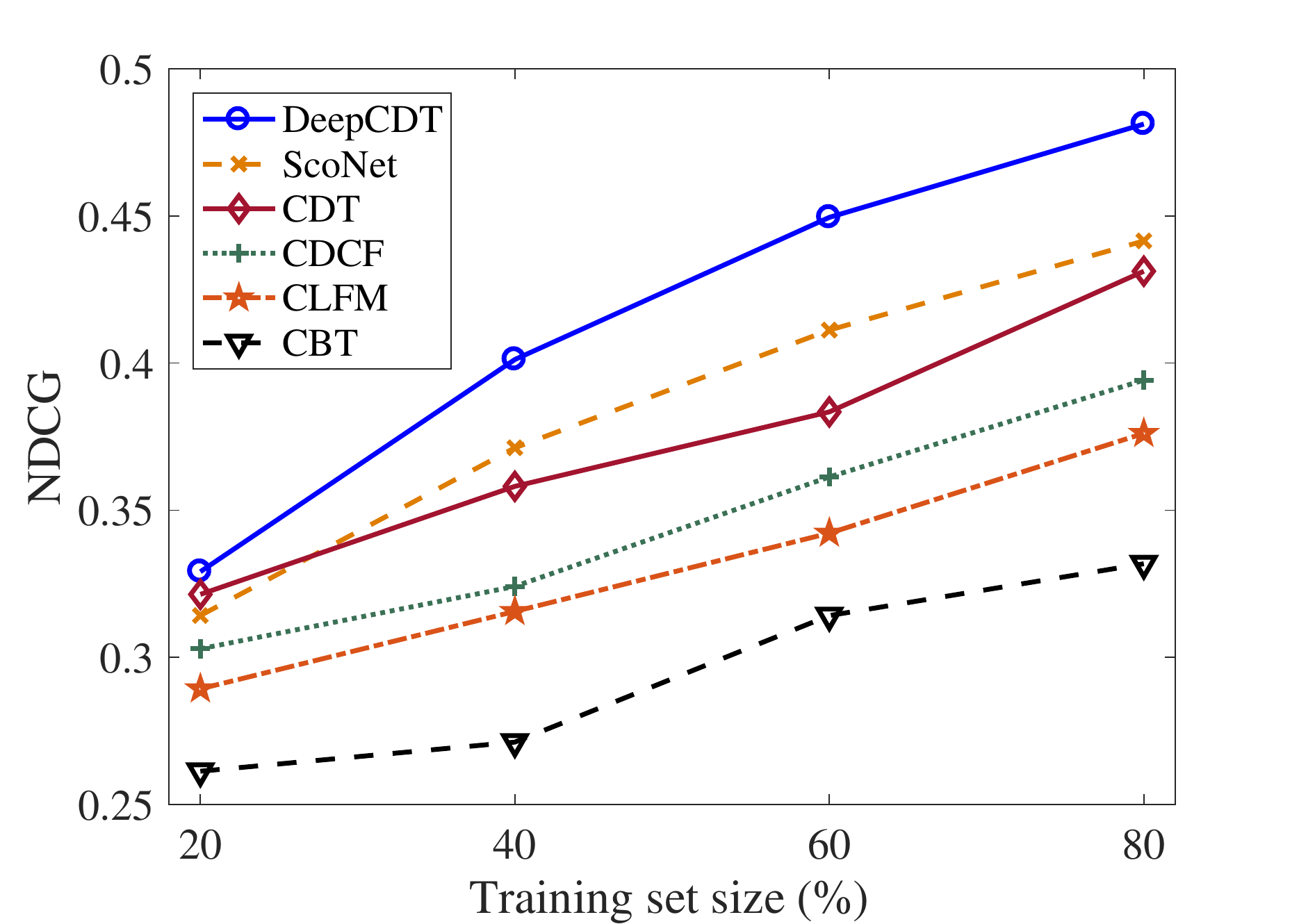}
\vspace{-0.3cm}
\caption{Effect on NDCG by varying the training set size for the cross-domain models in the ``Music'' domain.} \label{fig:perf}
\end{figure}

\bibliographystyle{unsrt}
\bibliography{transRec}

\end{document}

%% file: preamble.tex
\usepackage{cite}

\usepackage{amssymb,amsfonts,amsthm}
\usepackage[fleqn]{amsmath}

\usepackage{algorithmic}
\usepackage{graphicx}
\usepackage{textcomp}
\usepackage{xcolor}

\usepackage{algorithm}

\usepackage{graphicx}
\usepackage{caption}
\usepackage{subcaption}
\usepackage{url}

\usepackage{balance}

\usepackage{mdframed}

\def\BibTeX{{\rm B\kern-.05em{\sc i\kern-.025em b}\kern-.08em
    T\kern-.1667em\lower.7ex\hbox{E}\kern-.125emX}}

\usepackage{todonotes}

\DeclareCaptionLabelFormat{andtable}{#1~#2  \&  \tablename~\thetable}

%% file: transRec.bbl
\begin{thebibliography}{10}

\bibitem{Rendle10}
Steffen Rendle.
\newblock Factorization machines.
\newblock In {\em {ICDM}}, pages 995--1000, 2010.

\bibitem{AliannejadiRC17}
Mohammad Aliannejadi, Dimitrios Rafailidis, and Fabio Crestani.
\newblock Personalized keyword boosting for venue suggestion based on multiple
  lbsns.
\newblock In {\em Advances in Information Retrieval - 39th European Conference
  on {IR} Research, {ECIR} 2017, Aberdeen, UK, April 8-13, 2017}, pages
  291--303, 2017.

\bibitem{LI09}
Bin Li, Qiang Yang, and Xiangyang Xue.
\newblock Can movies and books collaborate? cross-domain collaborative
  filtering for sparsity reduction.
\newblock In {\em {IJCAI}}, pages 2052--2057, 2009.

\bibitem{Lon14}
Babak Loni, Yue Shi, Martha Larson, and Alan Hanjalic.
\newblock Cross-domain collaborative filtering with factorization machines.
\newblock In {\em {ECIR}}, pages 656--661, 2014.

\bibitem{RafailidisC16}
Dimitrios Rafailidis and Fabio Crestani.
\newblock Top-n recommendation via joint cross-domain user clustering and
  similarity learning.
\newblock In {\em Machine Learning and Knowledge Discovery in Databases -
  European Conference, {ECML} {PKDD} 2016, Riva del Garda, Italy, September
  19-23, 2016, Proceedings, Part {II}}, pages 426--441, 2016.

\bibitem{RafailidisC17}
Dimitrios Rafailidis and Fabio Crestani.
\newblock A collaborative ranking model for cross-domain recommendations.
\newblock In {\em Proceedings of the {ACM} Conference on Information and
  Knowledge Management, {CIKM}, Singapore, November 06 - 10, 2017}, pages
  2263--2266, 2017.

\bibitem{Hu18}
Guangneng Hu, Yu~Zhang, and Qiang Yang.
\newblock Conet: Collaborative cross networks for cross-domain recommendation.
\newblock In {\em {CIKM}}, pages 667--676, 2018.

\bibitem{GAO13}
Sheng Gao, Hao Luo, Da~Chen, Shantao Li, Patrick Gallinari, and Jun Guo.
\newblock Cross-domain recommendation via cluster-level latent factor model.
\newblock In {\em {ECML} {PKDD}}, pages 161--176, 2013.

\bibitem{HsiehYCLBE17}
Cheng{-}Kang Hsieh, Longqi Yang, Yin Cui, Tsung{-}Yi Lin, Serge~J. Belongie,
  and Deborah Estrin.
\newblock Collaborative metric learning.
\newblock In {\em {WWW}}, pages 193--201, 2017.

\bibitem{PasrichaM18}
Rajiv Pasricha and Julian McAuley.
\newblock Translation-based factorization machines for sequential
  recommendation.
\newblock In {\em RecSys}, pages 63--71, 2018.

\bibitem{HeKM17}
Ruining He, Wang{-}Cheng Kang, and Julian McAuley.
\newblock Translation-based recommendation.
\newblock In {\em RecSys}, pages 161--169, 2017.

\bibitem{LAP13}
Shenghua Gao, Ivor~Wai{-}Hung Tsang, and Liang{-}Tien Chia.
\newblock Laplacian sparse coding, hypergraph laplacian sparse coding, and
  applications.
\newblock {\em {IEEE} Trans. Pattern Anal. Mach. Intell.}, 35(1):92--104, 2013.

\bibitem{CREM11}
Paolo Cremonesi, Antonio Tripodi, and Roberto Turrin.
\newblock Cross-domain recommender systems.
\newblock In {\em {ICDMW}}, pages 496--503, 2011.

\bibitem{BPR}
Steffen Rendle, Christoph Freudenthaler, Zeno Gantner, and Lars Schmidt-Thieme.
\newblock Bpr: Bayesian personalized ranking from implicit feedback.
\newblock In {\em UAI}, pages 452--461, 2009.

\bibitem{ZhangDW16}
Weinan Zhang, Tianming Du, and Jun Wang.
\newblock Deep learning over multi-field categorical data - - {A} case study on
  user response prediction.
\newblock In {\em {ECIR}}, pages 45--57, 2016.

\bibitem{LeskovecAH07}
Jure Leskovec, Lada~A. Adamic, and Bernardo~A. Huberman.
\newblock The dynamics of viral marketing.
\newblock {\em {TWEB}}, 1(1):5, 2007.

\bibitem{Rec17}
Dimitrios Rafailidis and Fabio Crestani.
\newblock Learning to rank with trust and distrust in recommender systems.
\newblock In {\em RecSys}, pages 5--13, 2017.

\bibitem{AntarisRN14}
Stefanos Antaris, Dimitrios Rafailidis, and Alexandros Nanopoulos.
\newblock Link injection for boosting information spread in social networks.
\newblock {\em Social Netw. Analys. Mining}, 4(1):236, 2014.

\bibitem{RafailidisN15a}
Dimitrios Rafailidis and Alexandros Nanopoulos.
\newblock Repeat consumption recommendation based on users preference dynamics
  and side information.
\newblock In {\em Proceedings of the 24th International Conference on World
  Wide Web Companion, {WWW} 2015, Florence, Italy, May 18-22, 2015 - Companion
  Volume}, pages 99--100, 2015.

\end{thebibliography}
